\documentclass{article}
\usepackage{PRIMEarxiv}

\usepackage[numbers,sort&compress]{natbib}

\usepackage{graphicx}
\usepackage{verbatim}
\usepackage{color}
\usepackage{hyperref}
\usepackage{amsmath}
\usepackage{bm}
\usepackage{subfig}
\usepackage{multirow}
\usepackage[ruled,vlined,linesnumbered,noend]{algorithm2e}
\usepackage{combelow}
\usepackage{orcidlink}

\newlength{\commentWidth}
\newcommand{\atcp}[1]{\tcp*[r]{\makebox[\commentWidth]{#1\hfill}}}

\emergencystretch=\maxdimen
\title{CONTAIN: A Community-based Algorithm for Network Immunization}

\author{
  Elena-Simona Apostol$^{1,3}$\orcidlink{0000-0001-6397-4951}, 
  Özgur Coban$^{2}$, and
  Ciprian-Octavian Truic{\u{a}}$^2$\orcidlink{0000-0001-7292-4462}  \\
  $^1$National University of Science and Technology Politehnica Bucharest, 313 Independen\cb{t}ei, 060042, Bucharest, Romania \\
  $^2$Uppsala University, 1 Lägerhyddsvägen, 75105, Uppsala, Sweden \\
  $^3$Academy of Romanian Scientists, 3 Ilfov, Bucharest, Romania \\
  \texttt{elena.apostol@upb.ro, ozgur.coban.6046@student.uu.se, ciprian.truica@upb.ro}
}

\begin{document}

\maketitle              

\begin{abstract}
Network immunization is an automated task in the field of network analysis that involves protecting a network (modeled as a graph) from being infected by an undesired arbitrary diffusion.
In this article, we consider the spread of harmful content in social networks, and we propose \emph{CONTAIN}, a novel COmmuNiTy-based Algorithm for network ImmuNization.
Our solution uses the network information to
(1) detect harmful content spreaders, and
(2) generate partitions and rank them for immunization using the subgraphs induced by each spreader, i.e., employing \emph{CONTAIN}.
The experimental results obtained on real-world datasets show that \emph{CONTAIN} outperforms state-of-the-art solutions, i.e., NetShield and SparseShield, by immunizing the network in fewer iterations, thus, converging significantly faster than the state-of-the-art algorithms. 
We also compared our solution in terms of scalability with the state-of-the-art tree-based mitigation algorithm MCWDST, as well as with NetShield and SparseShield. We can conclude that our solution outperforms MCWDST and NetShield. 
\end{abstract}

\section{Introduction}

The adoption of advanced digital technologies has transformed and evolved social media, which in turn, enhanced the connectivity and awareness of our society. 
Along with these advancements, digitalization also created a favorable environment for the diffusion of misinformation~\cite{Ilie2021,Truica2022,Truica2022b,Truica2023} (i.e., the unintentional dissemination of false information), disinformation (i.e., the intentional dissemination of false information), and hate speech (i.e., the intentional dissemination of malicious content expressing hate and violence).
Thus, in the field of network analysis and graph mining, the current literature proposes novel approaches, strategies, and techniques for network immunization~\cite{zhang2015data,chen2015node,Petrescu2021,Zheng2018,Pham2020}.

Information diffusion tries to determine how the content produced by a node is spreading on a network. 
For example, in social network analysis, an information diffusion problem can be how the beliefs and biases of a politician influence followers and users~\cite{bakshy2012role}.
A related problem to the spread of information on a network, i.e., information diffusion, is influence maximization.
This optimization problem determines which nodes have the greatest influence in the network when dealing with the fast propagation and spread of information.
Thus, it focuses on selecting nodes that effectively accelerate the diffusion of information.

The complementary problem of influence maximization is network immunization.
Network immunization~\cite{weiss2013sir,saito2008prediction} strives to minimize the diffusion of harmful content within a network.
Similar to restraining the spread of viruses in a network modeled as a graph, network immunization marks and blocks a set of nodes, i.e., infected nodes, to prevent disseminating harmful content within a social network.

In the social network context, network immunization consists of strategies for \mbox{stopping} the spread of harmful content within a network structure.
The main goals of these strategies are to: 
\begin{itemize}
    \item[(1)] Detect the nodes that spread harmful content, i.e., spreaders, and 
    \item[(2)] Immunize the nodes within the network, i.e., followers, from such content.
\end{itemize}

Naturally, the problem of network immunization raises another one: whether the infection (i.e., the spreaders) is concentrated in clusters of nodes, specifically within communities of the network.
The likelihood of communities being created within networks, especially in social networks~\cite{porter2009communities}, provides a basis for this study.
Thus, in such scenarios where many communities exist in social networks, we aim to show that network immunization can be improved by employing a broad-scale approach at the network level, i.e.,  coarse granularity, and a more detailed approach at the community level, i.e., fine granularity.

In this article, the following research questions are investigated:
\begin{itemize}
    \item[(Q1)] Can a strategy centered around communities be employed for immunizing networks?
    \item[(Q2)] Does existing research in network analysis and sociology provide support for such an approach?
    \item[(Q3)] Can we improve the performance of network immunization over existing state-of-the-art methods?
\end{itemize}

The main objectives of this article are 
(1) to propose a new immunization strategy at the intersection of research in Network immunization, Community Structure Analysis, and Sociology and 
(2) to design a novel algorithm for Network Analysis of Social Networks where harmful content is prevalent.
Nevertheless, regardless of the specific context of our work, the results provided in this article can be generalized to a broader setting.
Thus, we propose \emph{CONTAIN} (COmmuNiTy-based Algorithm for network ImmuNization), a novel community-oriented network immunization algorithm for the social media environment based on immunizing communities. 
Using social context and network information, \emph{CONTAIN} detects the spreaders of harmful content, generates graph partitions, and ranks them. 
Our solution generates partitions based on the subgraphs induced by each spreader.
For our experiments, we use two real-life social media-based networks constructed from the FakeNewsNet~\cite{shu2018fakenewsnet} and Facebook Social Circles~\cite{leskovec2012learning} datasets, respectively.
We use the Louvain algorithm for detecting communities in graphs constructed from real-world social networks.
For classification, we train a model using the Random Forest algorithm.
We are aware that many works use more advanced Machine Learning and Deep Learning models for this task~\cite{Truica2022,Truica2022b,Truica2023,Truica2024,Truica2023MCWDST}, but the focus in this work is not fake news detection, but rather immunization.
We additionally use research results from Structural Holes and Dyadic Ties to justify our approach.
The experimental results on the two real-world datasets show that \emph{CONTAIN} outperforms state-of-the-art solutions, i.e., NetShield and SparseShield, by immunizing the network in fewer iterations, thus, converging significantly faster than the compared algorithms.
We also analyze the performance in terms of scalability on eight additional real-world graphs and compare it with two state-of-the-art solutions.

This paper is structured as follows.
In Section~\ref{sec:related_work}, we discuss the current literature and the state-of-the-art solutions on network immunization.
In Section~\ref{sec:methodology}, we present the methodology and the proposed novel network immunization algorithm \emph{CONTAIN}.
In Section~\ref{sec:experimental_results}, we perform the experimental evaluation and validation of \emph{CONTAIN} and compare the obtained results with two state-of-the-art network immunization algorithms, namely NetShield and SparseShield.
In Section~\ref{sec:discussion}, we discuss our findings and the limitations of this work.
Finally, in Section~\ref{sec:conclusions}, we summarize our findings and experimental results and hint at future work.

\section{Related Work}\label{sec:related_work}

In this section, we present state-of-the-art solutions for network immunization in social media and community-based network immunization.

\subsection{Network Immunization}

The network immunization solutions can be split into two classes: preemptive immunization and data-aware (or preventive) immunization.
Preemptive immunization algorithms leverage network features to proactively find a solution, whereas data-aware immunization techniques determine a solution concerning a specific diffusion seed.
DAVA (Data-Aware Vaccination Algorithm)~\cite{zhang2015data} is a data-aware immunization algorithm that takes a diffusion seed as input and then finds an immunization solution using a structure known as the dominator tree.
Yang et al.~\cite{Yang2018} propose a more time-efficient data-aware immunization solution that uses a dynamic node immunization heuristic that calculates, at each timestamp, the immunization gain of all nodes. It also employs a path-based influence model to identify the harmful nodes.
NetShield~\cite{chen2015node} is a preemptive immunization algorithm that computes a \textit{vulnerability} value equal to the dominant eigenvalue of the network and then builds a priority queue to immunize based on the budget.
SparseShield~\cite{Petrescu2021} is another preemptive immunization algorithm that takes as input the influence graph, the immunization budget, and the priority multiplier $\alpha$ to create the priority queue for immunization.
Afterward, it calculates the priority score for each node by utilizing the eigenvalues of the adjacency matrix.
For each node, a harmful score is computed and an ordered ranking list is created.
The ranking list is then employed to decide which nodes should be immunized based on an allocated immunization budget.

\subsection{Community-based Network Immunization}
Zheng and Pan~\cite{Zheng2018} consider the community structure of the network.
The solution detects the set of users where the misinformation originates and their community.
Besides reducing the influence of the harmful nodes to a given threshold, the heuristic tries to prevent influencing nodes that connect to other communities. 
Pham et al.~\cite{Pham2020} consider a more realistic scenario where misinformation from multiple topics can reach and affect the network at the same time.
The authors propose a Multiple Topics Linear Threshold model as an extension of the classical Linear Threshold model.
Ghalmane et al.~\cite{Ghalmane2019} use non-overlapping community structure to propose and investigate three node ranking strategies based on the (1) number of neighboring communities measure, (2) community hub‑bridge measure, and (3) weighted community hub‑bridge measure.
Wang et al.~\cite{Wang2020} propose a community-based immunization strategy consisting of three steps: (1) discover community structures using the modularity algorithm, (2) detect the candidate nodes for immunization, and (3) immunize the node using a memetic algorithm for selecting the seeds.
Experimental results show that the proposed solution manages to lower computational complexity, but does not converge.

\subsection{Community-based Fake News Network Immunization}
To the best of our knowledge, there is very little work regarding fake news detection using community detection.
Although~\cite{Galal2021} and~\cite{Simpson2022} adjacently tackle the same problem, these solutions are very different from our proposal, i.e., \emph{CONTAIN}.
CNMF~\cite{Galal2021} is a framework that utilizes a topic-based approach to extract communities and immunize the network.
The method proposed in~\cite{Simpson2022} tries to immunize the network by analyzing the diffusion, not communities.
These solutions do not quite approach the problem in a comparable manner to our proposed solution \emph{CONTAIN}.
\emph{CONTAIN} describes a highly modular algorithm that immediately lends itself to be incorporated into existing solutions and is further supported by research in other related fields, such as sociology.

\section{Methodology}\label{sec:methodology}

In this section, we present our novel community-based network immunization solution.
Firstly, we describe the proposed architecture.
Secondly, we present an overview of the classification models employed for detecting harmful content.
Finally, we introduce the proposed algorithm, \emph{CONTAIN}.

\subsection{Network Immunization Architecture}

\begin{figure*}[!ht]
\centering
\includegraphics[width=1\columnwidth]{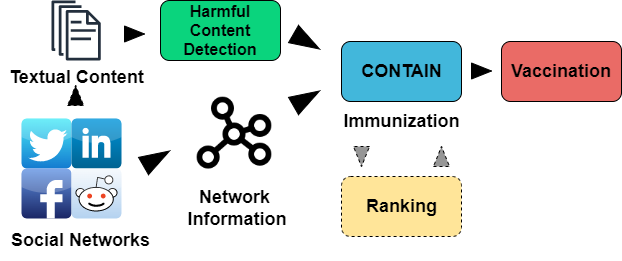}
\caption{The network immunization architecture}
\label{fig:arch}
\end{figure*}

Figure~\ref{fig:arch} presents the pipeline of our solution.
The Classification module is used to identify harmful nodes using the textual content.
We build multiple machine learning models to determine which one performs the best.
As the architecture is modular, we can choose the best-performing one for a given dataset or use all of them as an ensemble method.
The Immunization module employs \emph{CONTAIN}, our novel algorithm for network immunization.
The implementation of \emph{CONTAIN} relies on (1) detecting community structures and (2) leveraging the structural properties of the social network.
Thus, \emph{CONTAIN} is a community-oriented algorithm for network immunization.
The Immunization module contains a Ranking sub-module used for ranking the harmful node.
The Vaccination module allows the end-user to decide on the next steps. 
For example, it can delay harmful posts from the identified partitions ranked by the Immunization module.

\subsection{Harmful Nodes Detection}\label{ssec:hnd}
Before building a model for detecting the harmful content, we employ a text preprocessing pipeline with the following steps: 
(1) lowercase the text, 
(2) remove special characters, 
(3) remove numbers, 
(4) remove stop words,
(5) extract tokens. 
Moreover, because of grammatical variations and tense inflections, a corpus will contain numerous distinct forms of the same word.
In order to reduce the vocabulary size and eliminate such variations, we employ stemming, which transforms these words into a common base known as stem~\cite{manning2008introduction}.
We use Porter's algorithm~\cite{porter1980algorithm} for stemming.
Finally, we apply a preprocessing step to convert the data from raw text to a document-term matrix~\cite{Truica2016,Truica2020}.
We use TF-IDF (Term Frequency - Inverted Document Frequency) to build this matrix.
As TF-IDF combines the term frequency with the inverse document frequency, it manages to reflect both the importance of each word within the document and within the corpus of documents.
This assessment is also supported by the empirical experiments from~\citep{Truica2023}.

The classification module uses the following classification algorithms: 
(1) Random Forest,
(2) Extremely Randomized Trees,
(3) Stochastic Gradient Descent,
(4) Passive-Aggressive Algorithm,
(5) Logistic Regression, and
(6) Ridge Regression.
On a given dataset, we can select the best-performing model or we can use all these algorithms to create a voting ensemble model.
In our experiments, we choose the first option.

Each model is trained using K-fold cross-validation.
The training process involves the following steps:
\begin{itemize}
    \item[(1)] For each algorithm, we train a model using k-fold validation. At this step, we create a tuple $<$fold, algorithm$>$.
    \item[(2)] For each fold, we use grid search optimization to select the best hyperparameter for the machine learning algorithm. At this step, we expend the tuple and add the tested hyperparameters values, thus, obtaining tuples with the $<$fold,  algorithm, hyperparameters values$>$.
    \item[(3)] For each tuple $<$fold,  algorithm, hyperparameters values$>$, we record the accuracy, precision, and recall.
    \item[(4)] Finally, we compute the average and standard deviation for accuracy, precision, and recall for each pair $<$algorithm, hyperparameters values$>$ to obtain the model that has the best performance.
\end{itemize}

After detecting the harmful content, we then mark the nodes that spread such contact as harmful nodes.
These nodes are used as seeds for the immunization algorithms.

\subsection{Cross-Community Information Diffusion} 

Networks may display specific structural patterns that impact the spread of information.
These patterns are known as dyadic ties.
Dyadic ties can be classified into two observed variants: strong ties or weak ties~\cite{granovetter1973strength}.
Local cohesion and clustering within a network rely on strong ties, while information dispersion relies on weak ties.
This suggests the presence of weak ties connecting cliques to facilitate information flow between them, while strong ties should exist within each clique.
When two randomly selected nodes are connected in a network and each of them is a member of a different clique, information must flow through weak ties between the two cliques in order to go from one node to the other.
Granovetter~\cite{granovetter1973strength} demonstrates that weak ties play a crucial role in facilitating the diffusion of information across the entire network, while strong ties hinder diffusion within a community or subgraph due to their inherent local cohesion.
Therefore, the characteristics at the micro-level, i.e., fine granularity, of a network reveal the macro-level properties, i.e.,  coarse granularity.

Nevertheless, another structural pattern worth considering is Structural Holes.
Burt~\cite{burt2004structural} asserts and presents evidence indicating that there are structural holes between these groups due to homogeneity within groups.
These structural holes are filled by particular nodes referred to as brokers.
Brokers have the capability to participate in both groups, enabling them to influence the network at the coarse granularity level while existing at the fine granularity level.
Thus, brokers enable the transfer and diffusion of information between groups by connecting structural holes.

In summary, Granovetter~\cite{granovetter1973strength} highlights the significance of weak ties in the dissemination of information within a network, while Burt~\cite{burt2004structural} emphasizes the importance of structural holes in understanding the information flow within a network.
Consequently, Burt~\cite{burt2004structural} demonstrates the importance of structural holes when dealing with the information flow in a network.

Integrating sociological theories into our method serves as a foundation for understanding the behavior of communities and improving the immunization process.
To actually compute occupied structural holes of nodes, we employ the node constraint $c(v)$~\cite{Buskens2008} that relies on the edge weight $p_{uv}$ between nodes $u$ and $v$ (Equation~\eqref{eq:node_constant}). 
The constraint $c(v)$ quantifies the degree to which a node $v$ is linked to nodes within its neighborhood $N(v)$.
We employ the node constraint in our proposed solution to address the Research Question (Q2).

\begin{equation}\label{eq:node_constant}
c(v) = \sum_{u \in N(v)} (p_{uv} + \sum_{w \in N(u)} p_{uv}p_{wv})^2
\end{equation}

\subsection{Community Structures in Networks}\label{sec:communitystructures}

Community Structure Detection is based on the concept of modularity~\cite{newman2006modularity}.
For a network modeled as a graph $G=(V,E)$, it is defined as follows. 
Given graph $G=(V,E)$ with $n = |V|$ nodes and $m = |E|$ edges, where the set of nodes is divided into two subsets or groups $\varsigma_1$ and $\varsigma_2$, and its corresponding adjacency matrix $A$, where $A_{uv} = 1$ if there is an edge between the nodes $u$ and $v$ and $A_{uv}=0$ otherwise, we can define a membership variable for node $u$ as $s_u$ that denotes whether $u$ belongs to group $\varsigma_1$ (i.e., $s_v=1$ if $v \in \varsigma_1$) or to group $\varsigma_2$ (i.e., $s_v=-1$ if $v \in \varsigma_2$).
When edges are randomly placed between two nodes $u$ and $v$, the probable number of edges connecting them is indicated by $\varepsilon_{uv}$ (Equation~\eqref{eq:expected_edges}), where $d(u) = k_u$ and $d(v) = k_v$ are the node degree for $u$ and $v$, respectively.
Note, that the sum of all node degrees $\sum_{i \in V} d(i) = 2m - 1$ which in general is approximated to $2m$.
Finally, Equation~\eqref{eq:modularity}~\cite{newman2006modularity} presents the modularity $Q$ for graph $G$ which is obtained by summing over all pairs of nodes $u$ and $v$.

\begin{equation}\label{eq:expected_edges}
    \varepsilon_{uv} = \frac{d(u)d(v)}{\sum_{i \in V} d(i) } = \frac{k_u \cdot k_v}{2m -1} \approx \frac{k_u \cdot k_v}{2m}.    
\end{equation}

\begin{equation}\label{eq:modularity}
Q = \frac{1}{4m}\sum_{u,v} \left( A_{uv} - \varepsilon_{uv} \right) s_u s_v
\end{equation}

By creating a vector $\bm{s}^{T} = [ \dots s_i \dots ]$ that contains 
\textcolor{blue}{the collection of}  
all the membership variables $s_i$ ($i \in V$) and by redefining the modularity matrix as $\bm{B} = \{ b_{uv} | b_{uv} = A_{uv} - \varepsilon_{uv} \}$, we can rewrite the modularity as $Q = \frac{1}{4m} \bm{s}^T \bm{B} \bm{s}$.
By writing $\bm{s}$ as a linear combination of the normalized eigenvectors $\bm{\nu}_i$ of $\bm{B}$ with eigenvalues $\lambda_i$, the modularity becomes $Q = \frac{1}{4m} \sum_{i=1}^{|V|} (\bm{\nu}_i^T \cdot \bm{s})^2 \lambda_i$.
To partition the graph and apply the modularity formulation based on eigenvectors and eigenvalues, the algorithm follows these steps~\cite{newman2006modularity}:
\begin{itemize}
\item[(1)] Identify the leading eigenvector of the modularity matrix.
\item[(2)] Partition the nodes based on the sign of the vector elements.
\end{itemize}

The Louvain Community Detection method~\cite{blondel2008fast} is a novel approach that uses the greedy modularity technique to detect communities~\cite{Radulescu2020}.
To precisely calculate the change in modularity $\Delta Q$ (refer to Equation~\eqref{eq:deltaq}) for a given community $C$, where $m = |E|$ denotes the number of edges in graph $G=(V,E)$, $k_{v,in}$ is the sum of edge weights from vertex $v$ into community $C$, $k_v$ is the sum of weights in $E(v)$, $\sum_{tot}$ is the sum of weights in $E(C)$, and $\gamma$ is the resolution parameter.
Once the change in modularity $\Delta Q$ has converged to an optimum, a second phase commences by building another graph where the communities previously identified now function as nodes, and the first phase is iterated on this new network.

\begin{equation}\label{eq:deltaq}
    \Delta Q = \frac{k_{v,in}}{2m} - \gamma \frac{\sum_{tot} \cdot k_v}{2m^2}
\end{equation}

\subsection{CONTAIN}

We propose CONTAIN, a novel COmmuNiTy-based Algorithm for network ImmuNization -- to address the Research Question (Q1) --- by assuming that the network used the Independent Cascade Model~\cite{Saito2008}.
Formally, the problem we plan to solve is defined as follows. Given a graph $G$, a set of nodes that spread harmful content, and a budget, we aim to immunize the graph by detecting the communities where the harmful content is diffused.
The main components of the algorithm are the community structures embedded within the network.
As input, the algorithm receives a graph $G=(V, E)$, a seed set $S$, an immunization budget $k$, the resolution step value $\Delta \gamma$, and the resolution step $\gamma$.
In the case of social network immunization, the seed set $S$ contains the infected nodes, i.e., spreaders of malicious content. 
The immunization budget $k$ is the number of communities to immunize.
The resolution $\gamma$ determines how large the communities should be, i.e., $\gamma<1$ to select large communities, otherwise, select small communities.
This resolution is increased at each step with $\Delta \gamma$ until convergence.

\textit{CONTAIN} main steps are as follows:
\begin{itemize}
    \item[(1)] Retrieve the subgraph formed by each seed $s$ and their corresponding neighbors $N(s)$.
    \item[(2)] Create the subgraphs $G'$ and take note of any components by employing the node constraint measure $c(v)$ (Equation~\eqref{eq:node_constant}) to assess cross-community information diffusion.
    \item[(3)] Iteratively: 
    \begin{itemize}
        \item[(a)] Choose the partitions that have intersections with the components.
        \item[(b)] Generate partitions $C$ using the Louvain algorithm until a partition is found where the entire budget $k$ can be utilized.
    \end{itemize}
    \item[(4)] Provide as output a ranking indicating which partitions to prioritize for immunization.
\end{itemize}

In the pseudocode description of \emph{CONTAIN}, we presume that the composition of the subgraph always leads to a single component.
If this is not the situation, the task of subgraph composition can be subdivided, or the immunization strategy may need slight reformulation to take this into account.

Algorithm~\ref{alg:immunization} presents the proposed immunization algorithm.
The algorithm takes as input: 
\begin{itemize}
    \item[(1)] a graph $G=(V,E)$ for analysis,
    \item[(2)] a seed set $S$ containing identified malicious nodes by the detection model,
    \item[(3)] a community immunization budget $k$ that determines how many communities to immunize,
    \item[(4)] an initial resolution value $\gamma$ that is used by Louvain to determine the size of the recovered communities,
    \item[(5)] a resolution step value $\Delta \gamma$ that is used to increase the resolution value $\gamma$ in the case that the communities contain a small number of nodes.
\end{itemize}

The Louvain algorithm uses the resolution value $\gamma$ and the resolution step value $\Delta \gamma$ to decide the optimal size of communities for immunization.
The output of the algorithm is a ranked list of harmful communities arranged in descending order by their harmfulness score.
Concretely, \textit{CONTAIN} works as follows.
\begin{itemize}
    \item[(1)] Firstly, we extract the neighbors $N(s)$ for each seed $s \in S$ and compose the graph $G'$ (Lines~\ref{algo:line1} to~\ref{algo:line2}).
    \item[(2)] Compute the communities $C$ of the graph $G$ using the Louvain algorithm and the resolution $\gamma$ (Line~\ref{algo:line4}).
    \item[(3)] Check whether the partitions $C$ and the composed graph $G'$ meet the budget constraint $k$, and, if true, calculate and return a ranking of relevant communities (Lines~\ref{algo:line5} to \ref{algo:line7}).
    \item[(4)] Otherwise, the resolution $\gamma$ is incremented using the step value $\Delta \gamma$, and the algorithm continues with another iteration (Lines~\ref{algo:line8} to \ref{algo:line9}).

\end{itemize}

\begin{algorithm*}[!htbp]
\SetAlgoNoLine
\DontPrintSemicolon
\newcommand{\hrulealg}[0]{\vspace{1mm} \hrule \vspace{1mm}}

\SetKwInOut{Input}{Input}
\SetKwInOut{Output}{Output}
\SetKwFor{Loop}{Loop}{}{EndLoop}

\Input{$G$ - the graph for immunizing \newline 
$S$ - the list of harmful nodes, i.e., seeds \newline 
$k$ - the immunization budget \newline 
$\gamma$ - the resolution for communities' size \newline 
$\Delta \gamma$ -  the step value to increase the resolution
}
\Output{$r$ - the ranked list of harmful communities}

\hrulealg

\emph{$N(s) \leftarrow ExtractNeighbours(s \in S$)} \label{algo:line1} \atcp{extract the subgraph formed by each seed $s$}

\emph{$G' \leftarrow Compose(N(s))$} \label{algo:line2} \atcp{create the composed subgraph $G'$}

\Loop{}{\label{algo:line3}

    \emph{$C \leftarrow Louvain(G,\gamma)$} \label{algo:line4}  \atcp{extract communities using Louvain }

    \If{$|C \cap G'| \ge k$\atcp{verify if the budget $k$ constraint is meet}}{\label{algo:line5}
        \emph{$r \leftarrow Ranked(C \cap G')$}  \label{algo:line6} \atcp{rank harmful communities}
        \emph{\textbf{return} $r$} \label{algo:line7}    \atcp{return the ranked list}
    }
    \Else{\label{algo:line8} 
        \emph{$\gamma \leftarrow \gamma + \Delta \gamma$}  \label{algo:line9}  \atcp{increase the resolution}
    }
}

\caption{\emph{CONTAIN}: Community-based Algorithm for Network Immunization}
\label{alg:immunization}
\end{algorithm*}\DecMargin{1em}

The algorithm's time complexity is primarily influenced by Louvain, estimated at $O(n \log n)$, where $n=|V|$ represents the total number of nodes in the graph $G=(V, E)$.
The complexity for retrieving and composing subgraphs is constant, while for computing a rank the complexity is linear with the budget.
Furthermore, \emph{CONTAIN} implicitly employs structural holes, eliminating the need for explicit constraint computation.
Depending on the method used for ranking computation, it may reach $O(mn \log n)$, where $m$ is the number of nodes in the communities.
The description of rank computation is intentionally vague, as it can be implemented differently based on the context.
In our implementation of \emph{CONTAIN}, we utilize a ranking function $r(C)$ that uses as heuristic the ratio of harmful nodes $n_h$ within the total nodes of the community $n_C$, i.e., $r(C) = \frac{n_{h}}{n_{C}}$.
As the algorithm is predominantly concerned with communities, subgraphs, and groups, it inherently functions at a broad level of detail, i.e., coarse granularity level, promoting modularity and providing valuable insights into network immunization also at the fine granularity level by analyzing each community individually.

\section{Experimental Results}\label{sec:experimental_results}

In this section, we present the experimental results obtained for harmful content detection and network immunization with the novel \textit{CONTAIN} algorithm.
Additionally, we conduct a comparison of \textit{CONTAIN} with three advanced algorithms, namely SparseShield~\cite{Petrescu2021}, NetShield~\cite{chen2015node}, and MCWDST~\cite{Truica2023MCWDST}, as part of addressing the Research Question (Q3).
The obtained results are based on experiments performed on a virtual machine with the following hardware capabilities: 4GB of RAM and a single-core Intel(R) Core(TM) i5-4690K CPU @ 3.50GHz processor.
The operating system installed on this machine is Ubuntu 20.04 LTS.
The host for the virtual machine is a Lenovo System x3550 M4 server.

\subsection{Harmful Content Detection}

The dataset used for evaluation is the FakeNewsNet~\cite{shu2018fakenewsnet}, a dataset that contains both text and network information.
The textual information is labeled with its veracity, i.e., real or fake.

For this task, we train the models on textual content.
The models are implemented using Python and SciKit-Learn~\cite{pedregosa2011scikit}.
To determine what model manages to predict correctly the news articles' veracity, we employ 6 algorithms: (1) Random Forest,
(2) Extremely Randomized Trees,
(3) Stochastic Gradient Descent,
(4) Passive-Aggressive Algorithm,
(5) Logistic Regression, and
(6) Ridge Regression.

To assess the models, we employ accuracy, precision, and recall as evaluation metrics~\cite{Truica2017}.
Accuracy represents the ratio of correct predictions to total predictions.
Precision is the ratio of true positives to the sum of true positives and false positives. 
Recall is the ratio of true positives to the sum of true positives and false negatives.

\begin{table}[!ht]
\centering
\caption{Classification models evaluations (Note: \textbf{bold} text marks the overall best result)}
\label{tab:mlres}
\begin{tabular}{ lccc}
 \hline
 \textbf{Model} & \textbf{Accuracy} & \textbf{Precision} & \textbf{Recall}\\
 \hline
 Random Forest & $\mathbf{80.59 \pm 0.80}$ & $\mathbf{81.84 \pm  1.74}$  & $76.69 \pm 2.24 $  \\
 Extremely Randomized Trees & $77.57 \pm 1.02 $ & $78.47 \pm 1.55 $  & $76.46 \pm 1.94 $ \\
 Passive-Aggressive Classifier & $74.67 \pm 0.38 $ & $72.92 \pm 0.41 $  & $\mathbf{79.34 \pm 0.38}$ \\
 Stochastic Gradient Descent & $75.43 \pm 0.83 $ & $74.48 \pm 1.74 $  & $78.97 \pm 2.88 $ \\
 Logistic Regression & $71.43 \pm 0.00 $ & $71.38 \pm 0.00 $  & $72.63 \pm 0.00 $ \\
 Ridge Regression & $72.52 \pm 0.00 $ & $72.33 \pm 0.00 $  & $73.68 \pm 0.00 $ \\
 \hline
\end{tabular}
\end{table}

We train each model 10 times using K-Folds cross-validation.
Table~\ref{tab:mlres} presents the mean and standard deviation of each of the evaluation metrics.
Random Forest obtains the best results, with an accuracy of $80.59\% \pm 0.80\%$, a precision of $81.84\% \pm 1.74\%$, and a recall of $76.69\% \pm 2.24\%$.
To optimize the Random Forest model, we used a grid search approach that focused on the number of trees and the number of features to consider when looking for the best split.
Using 1\,400 trees and the square root as features estimator, it improves the accuracy to $81\%$.
We use the model obtained with Random Forest for the Harmful Content Detection module.

The nodes that spread harmful content are then marked as harmful nodes and are used as seeds for the immunization algorithm CONTAIN. Note, that these nodes are also used as seeds for the two state-of-the-art algorithms we use in our comparison, i.e., NetShield~\cite{chen2015node} and SparseShield~\cite{Petrescu2021}.

\subsection{CONTAIN Performance Analysis}

In this subsection, we perform an in-depth analysis of \emph{CONTAIN} and compare it to the state-of-the-art algorithms, i.e., NetShield~\cite{chen2015node} and SparseShield~\cite{Petrescu2021}.
Unfortunately, to the best of our knowledge, there are no other immunization algorithms that use a similar approach to ours for this comparison. 
Thus, we choose available algorithms that also perform network immunization given a seed of nodes.
It is important to acknowledge that comparing \emph{CONTAIN} with state-of-the-art solutions is not straightforward due to the distinct operational mechanisms of \emph{CONTAIN}, NetShield, and SparseShield, particularly regarding budget definition.
To facilitate meaningful comparisons, the task is slightly redefined for the state-of-the-art algorithms because \emph{CONTAINS} immunizes communities, rather than individual nodes.
Specifically, for both NetShield and SparseShield, rather than seeking a set of nodes to immunize within a given budget, the task is reformulated to determine a budget that achieves the immunization of the seed set.

The implementation of \emph{CONTAIN} is done in Python using the NetworkX~\cite{hagberg2008exploring} library.
In our comparison, we use the Python NetShield and SparseShield implementations from~\cite{logins2019experimental} and~\cite{Petrescu2021}, respectively.
The code is publicly available on GitHub at \url{https://github.com/DS4AI-UPB/CONTAIN}.

To evaluate \emph{CONTAIN}, we use the FakeNewsNet~\cite{shu2018fakenewsnet} and Facebook Social Circles ~\cite{leskovec2012learning} datasets.
Using these two datasets, we create two undirected graphs using the metadata provided with the dataset.
\ref{configs} presents FakeNewsNet Graph and Facebook Social Circles Graph properties, i.e., number of nodes and number of edges.
We select the seeds for the algorithms using the Harmful Content Detection module.

\begin{table}[!ht]
\centering
\caption{Graph properties}
\label{configs}
\begin{tabular}{ lrr}
 \hline 
 \textbf{Graph} & \multicolumn{1}{c}{\textbf{Nodes}} & \multicolumn{1}{c}{\textbf{Edges}}\\
 \hline
 FakeNewsNet Graph & 15\,257 & 459\,167 \\
 Facebook Social Circles Graph & 4\,039 & 88\,234 \\
 \hline
\end{tabular}
\end{table}

\begin{figure*}[!htp]
    \centering
    \subfloat[FakeNewsNet Graph\label{fig:g1}]{{\includegraphics[height=0.45\textheight]{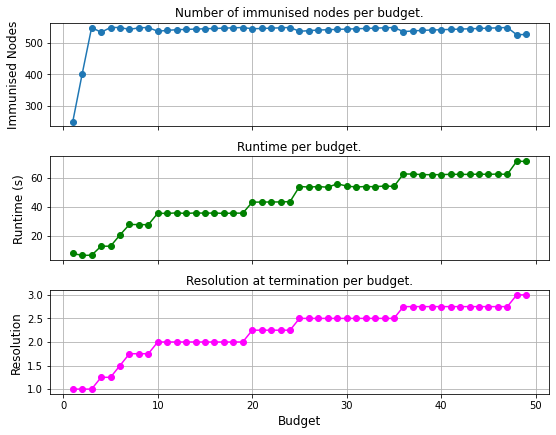}}}
    \hfill
    \subfloat[Facebook Social Circles Graph\label{fig:g2}]{{\includegraphics[height=0.45\textheight]{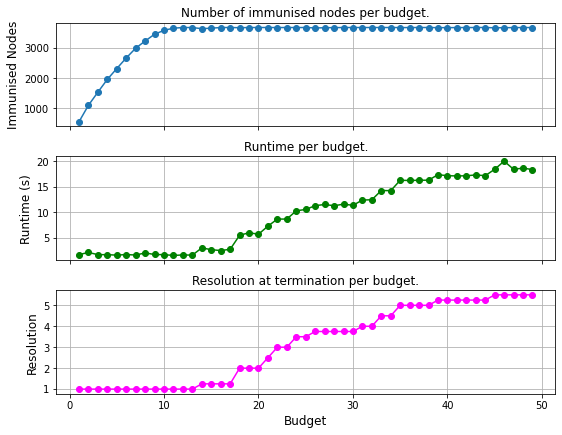}}}
    \caption{\emph{CONTAIN} performance}\label{fig:performance}
\end{figure*}

In Figures~\ref{fig:g1} and~\ref{fig:g2}, we observe how the number of nodes immunized actually converges, i.e., after a number of iterations the number of nodes does not change.
This observation remains consistent for both graphs, implying an inherent convergence property.
An informal proof of this concept is as follows: as the resolution approaches ``infinity'', Louvain identifies $n$ communities in a graph with $n$ nodes.
In simpler terms, the trivial communities consist of only one node, ensuring convergence.
As previously discussed when analyzing \emph{CONTAIN}'s complexity, Louvain plays a significant role in the algorithm's time complexity.
This becomes particularly evident when comparing the execution time with resolution since they seem to increase in direct correlation to each other.

As mentioned earlier, to conduct a comparison between \emph{CONTAIN} and the state-of-the-art algorithms, i.e., NetShield and SparseShield, it is necessary to adjust the task formulation.
Although \emph{CONTAIN} and the two state-of-the-art algorithms all take a budget and a seed of nodes as input (where the seeds are the nodes responsible for spreading harmful content), this reformulation is essential due to how the budget is defined.
For \emph{CONTAIN}, the budget $k$ is used to determine the number of communities to immunize. 
For NetShield and SparseShield, the initial definition of the budget is to determine the number of nodes to immunize.
In order to offer a better comparison between \emph{CONTAIN} and the state-of-the-art algorithms, for both NetShield and SparseShield, we are searching for a budget to immunize the seed set, i.e., the harmful nodes.

\begin{figure*}[!htp]
    \centering
    \subfloat[Runtime comparison. The red data points indicate the convergence for \emph{CONTAIN} and completion for NetShield/SparseShield.\label{comp1}]{{\includegraphics[height=0.45\textheight]{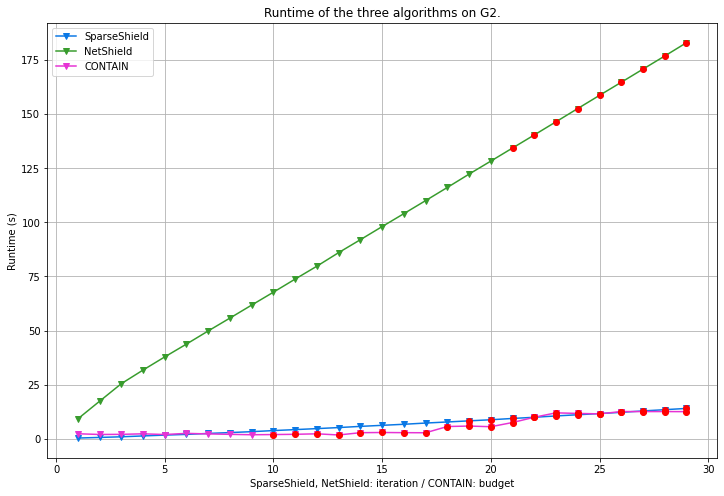}}}
    \hfill
    \subfloat[Number of nodes immunized. The red data points indicate the convergence for \emph{CONTAIN} and completion for NetShield/SparseShield.\label{comp2}]{{\includegraphics[height=0.45\textheight]{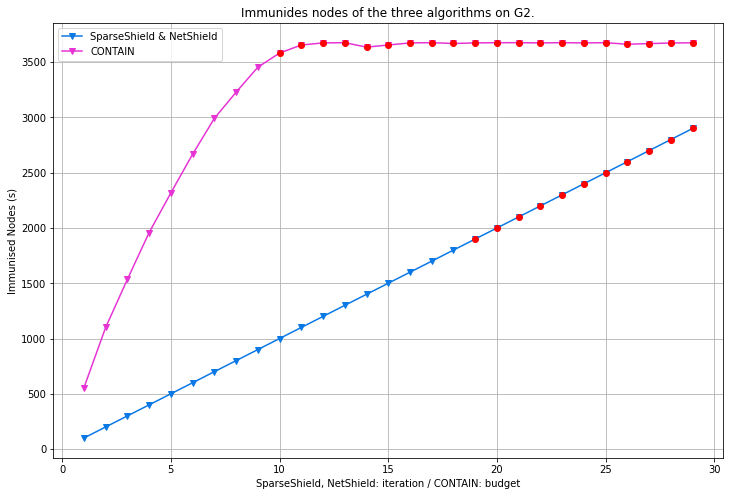}}}
    \caption{\emph{CONTAIN} vs. NetShield vs. SparseShield on Facebook Social Circles Graph}
\end{figure*}

The following set of experiments uses the graph extracted from Facebook Social Circles~\cite{leskovec2012learning} dataset.
Figure~\ref{comp1} presents the execution time of each algorithm.
We observe that \emph{CONTAIN} performs similarly to SparseShield.
Furthermore, the experimental results show that \emph{CONTAIN} outperforms NetShield.
Additionally, the memory demands of NetShield rendered it impractical to process the larger FakeNewsNet Graph, a limitation of NetShield also discussed by Petrescu et al.~\cite{Petrescu2021}.

Figure~\ref{comp2} presents the number of nodes immunized by \emph{CONTAIN}, NetShield, and SparseShield on the Facebook Social Circles Graph.
We notice that the significant operational distinction between the algorithms becomes evident.
Working at a higher level of abstraction does incur a cost in terms of precision compared to traditional methods. 
Nevertheless, this drawback can likely be alleviated with a refined ranking process.
Finally, we also take note of the marked data points as they show that \emph{CONTAIN} manages to immunize the network given a lower budget than NetShield and SparseShield.

\subsection{Scalability Analysis}

For this set of experiments, we use 8 real-world datasets presented in~\cite{Rozemberczki2019gemsec}.

For the first set of experiments, we randomly selected 10\% of the nodes for each graph to represent the seeds regardless of the algorithm.
For \emph{CONTAIN}, as it converges faster, we keep a budget equal to 10, while for SparseShield and NetShild we are still searching for a budget to immunize the network given the list of seeds.
Table~\ref{tab:scalability} presents the execution time in seconds and the number of immunized nodes for each algorithm.
As expected, the number of nodes immunized by NetShield and SparseShield is equal to the best possible budget we can determine empirically.
For \emph{CONTAIN}, we can observe that the number of immunized nodes is much higher, as it converges, while NetShield and SparseShield only immunize the seeds.
In terms of execution time, SparseShield is the fastest, while NetShield is the slowest.
\emph{CONTAIN} execution time is $\sim 3x$ to $\sim 5x$ slower than SparseShield w.r.t. the graph size.

\begin{table*}[!ht]
\centering
\caption{Scalability analysis \emph{CONTAIN} vs. NetShield vs. SparseShield}
\label{tab:scalability}
\resizebox{\columnwidth}{!}{
\begin{tabular}{lrrrrrrr}
\hline
\multirow{2}{*}{\textbf{Dataset}} & \multicolumn{1}{c}{\multirow{2}{*}{\textbf{Nodes}}} & \multicolumn{1}{c}{\multirow{2}{*}{\textbf{Edges}}} & \multicolumn{2}{c}{\textit{\textbf{CONTAIN}}} & \multicolumn{1}{c}{\multirow{2}{*}{\textbf{\begin{tabular}[c]{@{}c@{}}NetShield\\ Execution Time\end{tabular}}}} & \multicolumn{1}{c}{\multirow{2}{*}{\textbf{\begin{tabular}[c]{@{}c@{}}SparseShield\\ Execution Time\end{tabular}}}} & \multicolumn{1}{c}{\multirow{2}{*}{\textbf{\begin{tabular}[c]{@{}c@{}}NetShield/SparseShield\\ Immunized nodes\end{tabular}}}} \\
 & \multicolumn{1}{c}{} & \multicolumn{1}{c}{} & \multicolumn{1}{c}{\textbf{Execution Time}} & \multicolumn{1}{c}{\textbf{Immunized Nodes}} & \multicolumn{1}{c}{} & \multicolumn{1}{c}{} & \multicolumn{1}{c}{} \\
\hline
TV Shows & 3\,892 & 17\,262 & 1.12 & 1\,092 & 3.90 & 0.24 & 389 \\
Politician & 5\,908 & 41\,729 & 2.26 & 2\,765 & 15.54 & 0.52 & 590 \\
Government & 7\,057 & 89\,455 & 8.21 & 4\,100 & 26.72 & 1.54 & 705 \\
Public Figures & 11\,565 & 67\,114 & 6.23 & 6\,700 & 116.68 & 1.38 & 1\,156 \\
Athletes & 13\,866 & 86\,858 & 5.55 & 9\,052 & 202.57 & 1.81 & 1\,386 \\
Company & 14\,113 & 52\,310 & 3.82 & 5\,789 & 121.09 & 0.95 & 1\,411 \\
New Sites & 27\,917 & 206\,259 & 13.86 & 11\,582 & 1\,619.08 & 3.97 & 2\,791 \\
Artist & 50\,515 & 819\,306 & 158.34 & 38\,447 & 12\,637.32 & 32.05 & 5\,051
\\ \hline
\end{tabular}
}
\end{table*}

For the second set of experiments, we keep the number of immunized nodes fixed regardless of the algorithm.
We set the budget for CONTAIN equal to 5, i.e., CONTAIN must immunize up to 5 communities, and determine the number of nodes immunized.
We set the budget for NetShield and SparseShield equal to the number of nodes immunized by CONTAIN.
We also compare CONTAIN with the state-of-the-art tree-based mitigation algorithm MCWDST~\cite{Truica2023MCWDST}.
For the experiments with MCWDST, we set 
(1) the same seeds as CONTAIN, and 
(2) the budget equal to the number of nodes immunized by all the other algorithms.
Table~\ref{tab:scalability2} presents the results when all the algorithms immunize the same number of nodes. 
We observe that for this set of experiments, SparseShield is $\sim 2.5x$ to $\sim 4x$ faster than CONTAIN w.r.t. the graph size.
Although MCWDST performance is better than NetShiled, the execution time of SparseShield outperforms this algorithm.
We also note that CONTAIN is faster than MCWDST. 
This is in part due to 
(1) MCWDST transforms the graph into a tree representation, and 
(2) MCWDST performs more computations when determining the spread and gives scores to nodes to be immunized.

\begin{table*}[!ht]
\centering
\caption{Scalability analysis \emph{CONTAIN} vs. NetShield vs. SparseShield vs MCWDST for the same number of nodes to immunize}
\label{tab:scalability2}
\resizebox{\columnwidth}{!}{
\begin{tabular}{lrrrrrrrrr}
\hline
\multirow{2}{*}{\textbf{Dataset}} & \multirow{2}{*}{\textbf{Nodes}} & \multirow{2}{*}{\textbf{Edges}} & \multirow{2}{*}{\textbf{\begin{tabular}[c]{@{}c@{}}Immunized\\ Nodes\end{tabular}}} & \multirow{2}{*}{\textbf{\begin{tabular}[c]{@{}c@{}}CONTAIN\\ Execution Time\end{tabular}}} & \multirow{2}{*}{\textbf{\begin{tabular}[c]{@{}c@{}}NetShield\\ Execution Time\end{tabular}}} & \multirow{2}{*}{\textbf{\begin{tabular}[c]{@{}c@{}}SparseShield\\ Execution Time\end{tabular}}} & \multicolumn{2}{c}{\textbf{MCWDST Execution Time}}                                                                                                                                                       \\
                                  &                                 &                                 &                                                                                     &                                                                                            &                                                                                              &                                                                                                 & \textbf{\begin{tabular}[c]{@{}c@{}}Tree Construction\end{tabular}} & \multicolumn{1}{c}{\textbf{\begin{tabular}[c]{@{}c@{}}Immunization\end{tabular}}} \\ 
\hline
TV Shows       &  3\,892 &  17\,262 &     754  &  0.70 &       5.53 &  0.23 &   1.59 &   0.77 \\
Politician     &  5\,908 &  41\,729 &  1\,790  &  1.29 &      30.25 &  0.45 &  14.56 &   8.58 \\
Government     &  7\,057 &  89\,455 &  2\,857  &  3.54 &      29.76 &  0.91 &  51.24 &  15.06 \\
Public Figures & 11\,565 &  67\,114 &  5\,227  &  2.63 &     131.11 &  0.79 &  49.41 &  37.98 \\
Athletes       & 13\,866 &  86\,858 &  6\,108  &  2.34 &     237.79 &  1.03 &  87.46 &  65.26 \\
Company        & 14\,113 &  52\,310 &  4\,350  &  1.54 &     212.38 &  0.66 &  46.33 &  46.07 \\
New Sites      & 27\,917 & 206\,259 &  7\,585  &  7.31 &  1\,671.40 &  2.62 & 451.98 & 250.86 \\
Artist         & 50\,515 & 819\,306 & 30\,248  & 48.35 & 13\,173.27 & 13.73 & 5\,403.63 & 959.11
\\ \hline
\end{tabular}
}
\end{table*}

\section{Discussion and Limitations}\label{sec:discussion}

In this section, we ponder on the obtained results, placing them within a broader context to understand where the algorithm stands in the landscape of immunization methods.
The specific outcomes are carefully examined to assess the practical viability of \emph{CONTAIN}.
Following this, we delve into the reasoning behind the different implementation choices.

The algorithm's performance is predominantly determined by Louvain, influencing both time and memory complexity, serving as a bottleneck that requires refactoring for potential improvements.
Although on the tested graphs, convergence is reached at ${\sim}~15s$ and ${\sim}~1.5s$ for the larger FakeNewsNet Graph and for the smaller Facebook Social Circles Graph respectively, \emph{CONTAIN} dependence on Louvain might raise problems on graphs that contain millions of nodes and edges.
This can be mitigated using a Big Data Ecosystem as has been proposed by Truic{\u{a}} et al.~\cite{truica2018community}.
Furthermore, the scalability tests show that \emph{CONTAIN} has a small execution time that increases linearly with the size of the graph.

Operating at a coarse granularity inherently results in a loss of precision.
This implies that integrating a finely-tuned method, like SparseShield at a subsequent phase, could yield satisfactory results.
In the case of very large networks, iterating until convergence is identified might not be feasible.
In such situations, it could prove beneficial to initiate a randomized exploration of the network to identify feasible intervals before proceeding with an iterative search.
This strategy aligns with the analogous approach in training machine learning models.

\emph{CONTAIN} at its core, leverages the properties of social networks.
Therefore, when viewing the spread of malicious content as an intrinsic characteristic of engaging communities, the goal shifts to identifying harmful communities.
This stands in contrast to conventional immunization algorithms that primarily concentrate on identifying harmful nodes.
From a sociological perspective, focusing on a community as a cohesive entity is reasonable, suggesting that there is value in impeding the spread between communities.
This argument gains credibility, particularly when considering the insights of Granovetter~\cite{granovetter1973strength} and Burt~\cite{burt2004structural}.

In real-world social networks, there are community border nodes that flip between partitions.
By initially computing the subgraphs formed by the corresponding neighborhoods and subsequently recognizing the components of the resulting composition using the node constraint measure, we successfully identify communities in a deterministic fashion, even in situations where border nodes transition between different communities from one iteration to another.

In traditional immunization algorithms, budget determination involves calculating and detecting harmful nodes corresponding to the specified budget.
However, the objective of \emph{CONTAIN} varies from this traditional approach, as it aims to discover smaller communities that encompass harmful nodes.
Hence, using the traditional approaches, the budget criterion would be immediately fulfilled, since the basic community encompassing the entire graph consistently meets the budget requirement.
Consequently, our proposed approach involves calculating the budget for a specified number of communities, with a higher budget suggesting reduced collateral immunization.

\section{Conclusions}\label{sec:conclusions}

In this article, we propose \emph{CONTAIN} -- answering (Q1) --, a novel community-based algorithm for network immunization that: 
\begin{itemize}
    \item[(1)] Manages to immunize a network using a community-based approach, thus answering (Q1)
    \item[(2)] Incorporates sociological theories, i.e., node constraints, thus answering (Q2), and
    \item[(3)] Converges faster than state-of-the-art algorithms, thus answering (Q3).
\end{itemize}

The main contributions of this work are:
\begin{itemize}
    \item[(1)] \emph{CONTAIN}, a novel algorithm for performing community-oriented network immunization;
    \item[(2)] A modular architecture that allows the incorporation of other state-of-the-art methods for harmful content detection and network immunization;
    \item[(3)] A network immunization algorithm that operates integrates sociological theories and concepts.
\end{itemize}

The experimental results show that \textit{CONTAIN} performs similarly to the state-of-the-art algorithm SparseShield, while it outperforms NetShiels and MCWDST. 
Furthermore, \textit{CONTAIN} manages to converge much faster than both NetShield and SparseShield.
Our findings also confirm that the time complexity of the proposed algorithm is primarily influenced by the Louvain method.

In future work, we aim to develop a new modularity-based community detection algorithm to replace Louvain.
Furthermore, we also plan to adapt our approach for graph embeddings.

\section*{Contributions}
Elena Simona Apostol: Writing -- review \& editing, Writing -- original draft, Visualization, Validation, Supervision, Software, Resources, Project administration, Methodology, Investigation, Funding acquisition, Formal analysis, Data curation, Conceptualization 

\"Ozgur Coban: Visualization, Validation, Software, Resources, Methodology, Investigation, Formal analysis, Data curation, Conceptualization

Ciprian Octavian Truic{\u a}: Writing -- review \& editing, Writing -- original draft, Visualization, Validation, Supervision, Software, Resources, Project administration, Methodology, Investigation, Funding acquisition, Formal analysis, Data curation, Conceptualization

\section*{Declaration of competing interest / Conflict of interest}
The authors declare that they have no known competing financial interests or personal relationships that could have appeared to influence the work reported in this paper.

The authors declare no conflict of interest.

\section*{Acknowledgment}
This work is supported in part by
\begin{itemize}
    \item The National University of Science and Technology Politehnica Bucharest through the ``PubArt'' project. 
    \item The German Academic Exchange Service (DAAD) through the project ``iTracing: Automatic Misinformation Fact-Checking'' (DAAD grant no. 91809005).
    \item The Academy of Romanian Scientists through the funding of project ``SCAN-NEWS: Smart system for deteCting And mitigatiNg misinformation and fake news in social media'' (AO\cb{S}R-TEAMS-III).
\end{itemize}

The authors would like to thank the editors and the anonymous reviewers for providing insightful suggestions and comments
to improve the quality of this research paper.

\bibliographystyle{plainnat}
\bibliography{main}

\end{document}